\documentclass[preprint,prc,aps]{revtex4}
\usepackage{latexsym}
\everymath={\displaystyle}
\usepackage{rotating}
\def\1{\mbox{l\hspace{-0.53em}1}}

\usepackage{graphicx} 
\newcommand{\df}[2]{\ensuremath{ {\raise
1pt\hbox{$\displaystyle #1$}\over \raise -2pt \hbox{$\displaystyle
#2$}}}}
\begin{document}
\title{Can Y(4140) be a $c \bar c s \bar s$ tetraquark ?}

\author{Fl. Stancu\footnote{e-mail address: fstancu@ulg.ac.be}}
\affiliation{
 University of Li\`ege, Institute of Physics B5, Sart Tilman,
B-4000 Li\`ege 1, Belgium}

\date{\today}

\begin{abstract}
In this exploratory study the spectrum of tetraquarks of type 
$c \bar c s \bar s$ is calculated within a simple quark model with 
chromomagnetic interaction and effective quark masses extracted 
from meson and baryon spectra.
It is tempting to see if this spectrum can accommodate the resonance Y(4140), 
observed by the CDF collaboration, but not yet confirmed.
The results seem to favour the  J$^{PC}$ = 1$^{++}$ sector 
where the coupling to the VV channel is nearly as small as that of 
X(3872), when described  as a $c \bar c q \bar q$ tetraquark. This  suggests that 
Y(4140) could possibly be the strange partner of X(3872),
in a tetraquark interpretation. However the sector J$^{PC}$ = 0$^{++}$
cannot entirely be excluded. This work 
questions the practice of extracting 
effective quark masses containing spin independent contributions,
from mesons and baryons, to be used in multiquark systems as well.
\end{abstract}

\maketitle

\section{Introduction}

The CDF Collaboration \cite{Aaltonen:2009tz} has recently observed
a narrow structure in the $J/\psi \phi$ mass spectrum
of $B^+ \rightarrow J/\psi \phi K^+$ decays, which has been named 
Y(4140). Its mass and decay width are 
$M$ = 4143.0 $\pm$ 2.9(stat) $\pm$ 1.2(syst) MeV/c$^2$ and 
$\Gamma$ = 11.7$^{+8.3}_{-5.0}$(stat) $\pm$ 3.7(syst)  
MeV/c$^2$  respectively, which suggest that its structure does not fit 
conventional expectations for charmonium states. The CDF Collaboration 
expects that
the $J/\psi \phi$ final state, with positive C-parity and two J$^{PC}$ = 
1$^{--}$ vector mesons (VV), is a good candidate for an exotic meson search. 
This resonance is well above the threshold for open charm decay
$D^+_s D^-_s$ at 3936.68 MeV and a charmonium $c \bar c$
with this mass would  decay into an open charm pair
predominantly and have a small branching fraction into $J/\psi \phi$ 
\cite{Eichten:2007qx}. The mass of Y(4140) is  below the threshold 
of the decay channel $D^{*+}_s D^{*-}_s$ at 4224.6 MeV, and not
far above the $J/\psi \phi$ threshold at 4116.4 MeV.

More recently the Belle Collaboration reported preliminary results 
on Y(4140) \cite{Yuan:2009iu}. No significant signal was found 
but their efficiency is low for the mass of Y(4140). The upper
limit on the production rate 
$\mathcal{B}(B^+ \rightarrow Y(4140)K^+, Y(4140)\rightarrow J/\phi)$   
is 6 $\times ~10^{-6}$ at 90\% C.L. This upper limit is lower
than the  central value of the CDF measurement  
$(9 \pm 3.4 \pm 2.9) \times 10^{-6}$ \cite{Aaltonen:2009tz}
which is thus considered not to contradict the CDF measurement.

The Belle Collaboration also searched for Y(4140)   
in the $ J/\psi \phi$ mass spectrum of the two-photon process 
$\gamma \gamma \rightarrow J/\psi  \phi $ \cite{Yuan:2009iu}. 
Again, the efficiency
was low and no signal was reported. In exchange, evidence
was found for a new narrow structure at 4.35 MeV and width 13.3 MeV,
with a statistical significance of about $\sim 3.5\sigma$
in the $ J/\psi \phi $ mass spectrum. 
This resonance was named X(4350).

As such, the present situation allows a new opportunity  
to look for  exotics. The fashionable 
option of a $D^*_s \overline{D}^*_s$ molecule has been considered in Refs.
\cite{Liu:2009ei,Mahajan:2009pj,Ding:2009vd,Branz:2009yt}
and the QCD sum 
rules in Ref. \cite{Albuquerque:2009ak,Wang:2009ue,Zhang:2009st}.
where states with $J^{PC} = 0^{++}$ or $2^{++}$ are favoured.
Let us note however
that the Belle Collaboration measurement of a two-photon partial
width  difavours the scenario of Y(4140) to be a
$D^*_s \overline{D}^*_s$ molecule  with $J^{PC} = 0^{++}$
or $2^{++}$ \ \cite{Yuan:2009iu}.

Prior to the observation of Y(4140) by the CDF Collaboration, predictions 
for tetraquarks   $c \bar c s \bar s$  seen as diquark-antidiquark systems
with various $J^{PC}$ 
were made 
in a simple non-relativistic model including
$\ell$ = 0 and 1 partial waves 
in Ref. \cite{Drenska:2009cd} and in a relativistic framework based
on the quasipotential approach in
Ref. \cite{Ebert:2005nc}. In the latter, states with $0^{++}$ and
$1^{+\pm}$ acquired masses in the range 4.1 - 4.2 MeV. 

We should also mention that the resonance Y(4140) was studied as the 
second radial excitation of the P-wave charmonium $\chi_{cJ}^{\prime\prime}$
($J$ = 0 and 1), looking at the hidden charm decay mode. The conclusion
was that such a description is problematic \cite{Liu:2009iw}.

Deciphering the nature of Y(4140), if confirmed in the future,
(presently the B-factories have a poor acceptance
for $B \rightarrow K J/\psi \phi$ in the desired range \cite{Olsen:2009ys}),
is a new challenge. 
Thus it is legitimate  to consider the tetraquark interpretation
without correlated quarks or antiquarks, 
and try to find out if the Y(4140) fits into
the spectrum of the $c \bar c s \bar s$ system.  Most important,
we search for the decay pattern
given by this possible structure. For simplicity, we use the model of
Ref. \cite{Hogaasen:2005jv} which successfully describes the X(3872)
as a $c \bar c q \bar q$ tetraquark. In Ref. \cite{Hogaasen:2005jv}
it was shown that X(3872) can be interpreted as 
an eigenstate of the chromomagnetic interaction, where the lowest 
1$^{++}$ has a dominant octet-octet component (0.9997)
and a very small singlet-singlet component (0.026)
which explains why this state decays with a very small width
into $J/\psi + \rho$ or 
$J/\psi + \omega$, in agreement with the experimental
value for the total width $\Gamma <$ 2.3 MeV of X(3872)\ \cite{Choi:2003ue},
and that the $J/\psi$ + pseudoscalar channel is absent.
As  Y(4140) is seen to be narrow and decays into two vector mesons 
we wonder whether or not  the same mechanism can give an
explanation of its small width, about 5 times larger than that
of X(3872), and similar to that of X(4350), but considerably narrower than the 
decay width of every other X,Y or Z resonances. 

The paper is organized as follows. In Sec. \ref{HAM}   we introduce
the quark model used in this study. In Sec. \ref{BS} we recall 
the basis states in the direct meson-meson channel with emphasis
on the charge conjugation quantum number. In Sec.  \ref{ME} we present
the matrix elements of the Hamiltonian \cite{Hogaasen:2005jv} 
for $J^{PC} = 0^{++}, 1^{++},
1^{+-}$ and  $2^{++}$ states. In Sec. \ref{SPECTRUM} we show the calculated 
spectrum and discuss its features. The last section is devoted to conclusions.
In Appendix \ref{appa} we derive the orthogonal transformation 
from the direct   meson-meson channel to the exchange meson-meson channel
for states $0^{++}$, in Appendix \ref{appb} for states with
$1^{++}$ and $1^{+-}$ and in Appendix \ref{appc} for states with $2^{++}$.
Appendix D is devoted to an attempt to dynamically derive effective masses
in a standard constituent quark model in order to justify the simplicity  
of the present study and enlighten the choice of effective masses.  
\section{The model}\label{HAM}

This is an exploratory study, based
on the simple model of Ref. \cite{Hogaasen:2005jv}
which can reveal the  
basic features of the $c \bar c s \bar s$ tetraquark, especially the
structure of the wave functions.  In the next section 
we introduce the relevant 
basis states in the color-spin space, including both the 
singlet-singlet channels and the octet-octet, simply called hidden color 
channels. 
There are no correlated quarks or
diquarks, as in Ref. \cite{Drenska:2009cd}, for example. 

Accordingly, the mass of a tetraquark is given by the
expectation value of the effective Hamiltonian \cite{Hogaasen:2005jv}
\begin{equation}\label{eq:Mcm}
H =\sum_i m_i +  H_{\mathrm{CM}} ,
\end{equation}
where
\begin{equation}\label{eq:Hcm}
H_{\mathrm{CM}} = - \sum_{i,j} C_{ij}\,
~\lambda^c_{i} \cdot \lambda^c_{j}\,\vec{\sigma}_i \cdot 
\vec{\sigma_j}~.
\end{equation}
The first term in Eq. (\ref{eq:Mcm}) contains the effective masses $m_i$ 
as parameters. The constants $ C_{ij}$
represent integrals in the orbital space of some unspecified 
radial forms of the chromomagnetic part of the one gluon-exchange 
interaction potential and of the wave functions.

A warning should be given to the way of determining the effective masses $m_i$
to be used for  multiquark systems.
Besides the kinetic energy contribution, they incorporate the effect of 
a Coulomb-like term and of the 
confinement, the latter 
still being an open problem 
\cite{Vijande:2007ix}. Thus, in principle, they 
cannot be directly extracted from meson or baryon spectra
as discussed in Appendix D. Lack of better
knowledge we however 
use the compromise proposed in Ref. \cite{Hogaasen:2005jv}
\begin{equation}\label{eq:mass}
\begin{array}{lll}
 m_{c}=1550\,\mathrm{MeV},&
m_{s}=590\,\mathrm{MeV},
\end{array}\end{equation}
but due to the arbitrariness in the choice of effective masses of quarks, 
precise estimates of the absolute values of tetraquark masses 
is difficult to make. One can have an approximate idea about the range 
where the spectrum should be located. But a shift of the whole 
spectrum is justified and sometimes even performed, like in the
popular work of Maiani et al.  \cite{Maiani:2004vq}, which deals
with diquarks, where the arbitrariness in mass is even larger.

However,  the relative distances between the eigenstates obtained  
from the chromomagnetic Hamiltonian  (\ref{eq:Hcm}) and the structure 
of its eigenstates do not depend on the effective masses,
which is important for exploring the strong decay properties. 

The  parameters $C_{ij}$ have been taken from  Ref.\  \cite{Buccella:2006fn} 
where a more complete list, containing also parameters needed in this work,
is given. The required values are 
\begin{equation}\label{eq:par}
\begin{array}{lll}
%
C_{cs}=5.0\,\mathrm{MeV}, &C_{c\overline{c}}=5.5\,\mathrm{MeV},\\
C_{c \bar s}=6.7\,\mathrm{MeV}, &C_{s \bar s} = 8.6\,\mathrm{MeV}. 
\end{array}\end{equation}
We should mention that the above parameters were extracted from a
global fit to meson and baryon ground states. For some mesons into which 
Y(4140) can decay  in Table \ref{pdgmass} we compare 
the experimental masses of PDG \cite{Amsler:2008zzb}
with the theoretical values 
obtained from the two-body version of (\ref{eq:Mcm}) and (\ref{eq:Hcm})
in the parametrization (\ref{eq:par}) which is
\begin{equation}\label{meson}
m_{q \bar q} = m_q +  m_q - \langle \lambda^c_1 \cdot \lambda^c_2 \rangle
\langle \vec{\sigma_1} \cdot \vec{\sigma_2} \rangle  C_{q \bar q}
\end{equation}
where q stands here for any light or heavy quark.
\begin{table} 
\caption{Theoretical and experimental meson masses in MeV}
\label{pdgmass}
\renewcommand{\arraystretch}{1.8}
 \begin{tabular}{ccccccc}
\hline 
Meson    &\hspace{1cm}  & $J^{PC}$ &\hspace{1cm} & Theory  & \hspace{1cm} & Exp  \\  
\hline
$J/\psi$ & \hspace{1cm} & $1^{--}$ & \hspace{1cm} & 3121.3  & \hspace{1cm} & 3096.9 \\
$\phi$   & \hspace{1cm} & $1^{--}$ & \hspace{1cm} & 1225.9 & \hspace{1cm} & 1019.5 \\
$D_s $   & \hspace{1cm} & $0^{-?}$ & \hspace{1cm} & 2032.0 & \hspace{1cm} & 1968.5 \\
$D^*_s$  & \hspace{1cm} & $1^{-?}$ & \hspace{1cm} & 2175.7 & \hspace{1cm} & 2112.3 \\
\hline
\end{tabular}
\end{table}
From Table \ref{pdgmass} one can see that the two-body Hamiltonian (\ref{meson})
with the masses effective (\ref{eq:mass}) systematically 
overestimates the meson masses.
Therefore the threshold energies of the channels $J/\Psi \phi$,
$D_s {\overline D}_s$, $D^*_s {\overline D}^*_s$,
and $D^*_s {\overline D}^*_s$, are considerably
overestimated. Due to this discrepancy it is meaningless to compare 
the tetraquark states with the theoretical threshold.   
This work questions 
the practice of using identical effective masses in both ordinary and exotic
multiquarks. In such a case we would return us to the schematic treatment 
of the never observed "stable"
H-dibaryon \cite{Jaffe:1976yi} 
predicted to be strongly bound by the chromomagnetic 
interaction.
We do not intend to make a fine tuning of the effective masses. 
We are mostly interested 
in the structure of the tetraquark wave functions which essentially depends 
on the hyperfine interaction.
We shall compare the calculated spectrum to the experimental thresholds.
In Appendix D we give a simple proof that one cannot use the same 
effective masses both in mesons and tetraquarks. 

In the following, an important parameter in this study is the
difference  between the values of $C_{c \bar s}$ and $C_{cs}$.
In fact we shall see that the replacement of the light quarks $q = u, d$
with the strange quark $s$ does not much modify the structure of the  
$c \bar c s \bar s$ with respect of that of $c \bar c q \bar q$.
 
\section{The basis states}\label{BS}

Here we use a basis vectors relevant for understanding the decay properties
of tetraquarks.  The total wave function of a tetraquark is a linear 
combination of these vectors.
We suppose that particles 1 and 2 are quarks and particles 3 and 4 
antiquarks, see Fig. 1.  In principle the basis vectors should contain
the orbital, color, flavor and spin degrees of freedom such as to account 
for the Pauli principle. But, as we consider $\ell = 0$ states 
the orbital part is symmetric and anyhow irrelevant for the effective
Hamiltonian described in the previous section. Moreover, 
as the flavor operators do not explicitly appear in
the Hamiltonian, the flavor part does not need to be specified. 
A detailed description of the three distinct bases  corresponding to 
the three choices of internal coordinates shown in Fig. 1 is presented in 
Refs.\ \cite{Brink:1994ic,Brink:1998as}.
It was found that the inclusion of meson-meson channels 
accelerate the convergence, for example in $cc \bar q \bar q$ tetraquarks 
\cite{Janc:2004qn}.

\begin{figure}\label{fig1}
\begin{center}
\includegraphics*[width=10.0cm,keepaspectratio]{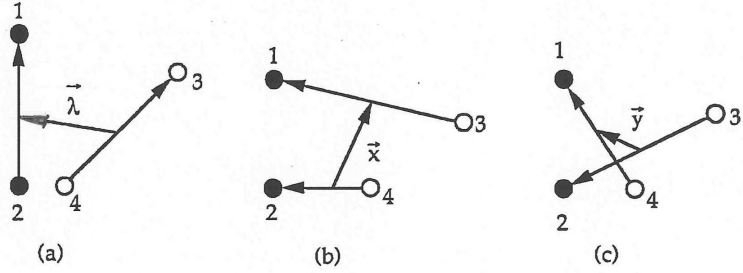}
\end{center} 
\caption{Three independent relative coordinate systems. Solid and 
open circles represent quarks and antiquarks respectively: (a)
diquark-antidiquark channel, (b) direct meson-meson channel, (c) exchange
meson-meson channel. }
\end{figure}

We remind that in the color space there are three 
distinct bases: 
~$ a)~|\overline{3}_{12} 3_{34} \rangle, ~ |6_{12} \overline{6}_{34} \rangle $,
~$ b)~|1_{13} 1_{24} \rangle, ~ |8_{13} 8_{24} \rangle $ ,
and 
~$ c)~|1_{14} 1_{23} \rangle, ~ |8_{14} 8_{23} \rangle$, 
associated to the three distinct internal coordinate systems 
shown in Fig. 1. 
The 3 and $\overline{3}$ are antisymmetric and 6 and $\overline{6}$
are symmetric under interchange of quarks and antiquarks respectively.
This basis is convenient for diquark-antidiquark models, 
where usually the color space is truncated to contain only 
$|\overline{3}_{12} 3_{34} \rangle$ states \cite{Maiani:2004vq}.
This reduces 
each J$^{PC}$ spectrum  to twice less states than allowed 
by the Pauli principle \cite{Stancu:2006st}
and influences the tetraquark properties.
The sets $b)$  and $c)$ 
contain a singlet-singlet color
and an octet-octet color state. The amplitude of the latter
vanishes asymptotically, when the mesons, into which a tetraquark 
decays, separate. These
are called \emph{hidden color} states by analogy to states
which appear in  the nucleon-nucleon problem, defined as a six-quark system
\ \cite{HARVEY}.
The contribution of hidden color states to the 
binding energy of light tetraquarks has been calculated explicitly in 
Ref.  \cite{Brink:1994ic}.
Below we shall point out their  role in the description 
of the of $c \bar c s \bar s$ tetraquarks.  
The situation is similar to the interpretation of the X(3872) resonance
as a $c \bar c q \bar q$ tetraquark in Ref. \cite{Hogaasen:2005jv},
where its small width has been explained as due to a tiny 
J/$\psi$ + $\rho$ or J/$\psi$ + $\omega $ component in the wave function 
of the 1$^{++}$ tetraquark state.


As the quarks and antiquarks are spin 1/2 particles the total 
spin of a tetraquark can be $S = 0$ , $S = 1$  
or $S = 2$. 

For $S = 0$ there are two independent basis states (two Young tableaux)
for each channel.
The spin states associated to the three distinct internal coordinates 
depicted in Fig. 1 are:
~$a)~ |S_{12}S_{34} \rangle,~~|\vec{A}_{12} \cdot \vec{A}_{34} \rangle$,
~$b)~ |P_{13} P_{24} \rangle,~~ |(V_{13} V_{24})_0 \rangle$ , 
~$c)~ |P_{14} P_{23} \rangle,~~ |(V_{14} V_{23})_0 \rangle$, 
respectively, where $S$ stands for scalar,  $A$ for axial and $P$ 
and $V$ for pseudoscalar and vector 
subsystems  and the lower index 0 indicates the total spin.
The relation between the three different bases can be found in 
Ref. \ \cite{Brink:1998as}.

For $S = 1$ there are three independent spin states, corresponding to three
distinct Young tableaux. 
Presently we are interested into those corresponding to Fig. 1b,
named the direct meson-meson channel. In this channel 
we remind that the basis vectors are
\cite{Brink:1998as}
\begin{equation}\label{directcolor1}
|(P_{13} V_{24})_1 \rangle, ~~~~|(V_{13} P_{24})_1 \rangle, 
~~~|(V_{13} V_{24})_1 \rangle.
\end{equation}
As above, the lower index indicates the total spin 1.

In this case the charge conjugation operator is related to
permutation properties of the basis vectors in a simple way.
Under the transposition (13)  manifestly one has
\begin{equation}
    (13) |P_{13} \rangle = - |P_{13} \rangle,
~~~ (13) |V_{13} \rangle = + |V_{13} \rangle,
\end{equation}
and similarly for the transposition (24) 
\begin{equation}
    (24) |P_{24} \rangle = - |P_{24} \rangle,
~~~ (24) |V_{24} \rangle = + |V_{24} \rangle.
\end{equation}

The case $S = 2$ is trivial. There is a single basis state 
\begin{equation}\label{SPIN2}
\chi^S = |(V_{13} V_{24})_2 \rangle , 
\end{equation}
which is symmetric under any permutation of quarks.

From Ref. \cite{Stancu:1991rc} Ch. 10,  one
can see that the permutation (13)(24) leaves invariant the 
color basis vectors $|1_{13} 1_{24} \rangle$ and $|8_{13} 8_{24} \rangle $. 
Then, with the identification 1 = $c$, 2 = $s$, 3 = $\overline c$ and 
4 = $\overline s$ the permutation (13)(24) is equivalent to the
charge conjugation operator \cite{Stancu:2008zh}. Thus all basis states 
introduced below have a definite charge conjugation, which is easy to identify.

\section{Matrix elements}\label{ME}

For a ground state tetraquark the possible states are
$J^{PC} = 0^{++}, 1^{++}, 1^{+-}$ and $2^{++}$.
In the direct meson-meson channel, in  each case a basis can be built
with the quark-antiquark pairs (1,3) and (2,4) as subsystems, where each
subsystem has a well defined color state, a singlet-singlet or an octet-octet. 
This arrangement is convenient to describe hidden charm 
$J/\psi $ + \emph{light meson}
or $\eta_c$ + \emph{light meson} channels, the light meson quantum
numbers being consistent with $J^{PC}$.
The other quark-antiquark pairs,
(1,4) and (2,3) describe open charm
meson channels, here called exchange channels (see Fig. 1c)
 as \emph{e. g.} $D_s {\overline D}_s$,
$D_s {\overline D}^*_s$ or
$D^*_s  {\overline D}^*_s$.   
One can fix a basis in terms of the problem one looks at, but
for convenience, in the calculations one can
pass from one basis to another by an orthogonal transformation. 
In this study the adequate basis is that related to the direct
meson-meson channel, depicted in Fig. 1b. 
The orthogonal transformations from the direct to the exchange 
meson-meson channel for $J^{PC}$ = $0^{++}$ and $1^{++}$
are given in Appendices \ref{appa} and \ref{appb} respectively.

The matrix elements introduced below appeared in the Proceedings
\cite{Stancu:2006st}. For the reader's convenience we present
them here again. They correspond to the scalar, axial and tensor
tetraquarks introduced above. Later on, the authors of 
Ref. \cite{Buccella:2006fn} calculated the matrix elements of the
chromomagnetic interaction (\ref{eq:Hcm}) in a basis corresponding to
Fig. 1a. Although the spectrum is the same,
one cannot distinguish between 
charge conjugation $C = 1$ and $C = -1 $ because 
in that basis $J^P = 1^+$ states do not have a definite charge conjugation. 
To identify $C$ one must return to our basis.   
Therefore we found it convenient to use our basis which can give direct
information to experimentalists. 

For $J^{PC} = 0^{++}$ the basis constructed from products of color
and spin states 
associated to Fig. 1b are
\begin{eqnarray}\label{eq:betai}
\hskip -10pt& \psi^1_{0^{++}} = 
| 1_{13} 1_{24} P_{13} P_{24} \rangle,\
&  \psi^2_{0^{++}} =
| 1_{13} 1_{24} (V_{13} V_{24})_0 \rangle, \nonumber\\
\hskip -10pt& \psi^3_{0^{++}} =
| 8_{13} 8_{24} P_{13} P_{24} \rangle,\ 
&\psi^4_{0^{++}} =
| 8_{13} 8_{24}  (V_{13} V_{24})_0 \rangle .
%
%
\end{eqnarray}
The chromomagnetic interaction Hamiltonian with minus sign, -$H_{\mathrm{CM}}$,
acting on this basis leads to the following symmetric matrix
\\ 
\vskip -20pt
\begin{equation}\left[
\renewcommand{\arraystretch}{2.0}
\begin {tabular}{cccccc}
%
$16(C_{13}+C_{24})$&0&0&$ 8 \sqrt{\df{2}{3}}(C_{12}+C_{23})$\\
%
 &$-
\df{16}{3}(C_{13}+C_{24})$&$ ~~-8 \sqrt{\df{2}{3}}(C_{12}+C_{23})$
&$\df{16 \sqrt2}{3}(C_{23}-C_{12})$\\
%
%
 & &$-2(C_{13}+C_{24})$ & $\df{4}{\sqrt3}(2C_{12}-7C_{23})$\\
%
%
 & & & $\df{16}{3}C_{12}+\df{56}{3}C_{23}+\df{2}{3}(C_{13}+C_{24})$
\\
\end{tabular}\right]
\label{0++}
\end{equation}


For $J^P=1^{++}$ there are two linearly independent basis 
vectors built as products of color 
and the third spin state of Eq. 
(\ref{directcolor1}) .
%
\begin{eqnarray}\label{eq:alphai}
%
\hskip -10pt& \psi^1_{1^{++}} =
| 1_{13} 1_{24}~ (V_{13} V_{24})_1 \rangle, \  
& \psi^2_{1^{++}} =
| 8_{13} 8_{24}~ (V_{13} V_{24})_1 \rangle.
\end{eqnarray}
The matrix associated to the chromomagnetic interaction -$H_{\mathrm{CM}}$ is
\\
\vskip -20pt
\begin{equation}\left[
\renewcommand{\arraystretch}{2.0}
\begin {tabular}{cc}
%
$ -\df{16}{3}(C_{13}+C_{24})$ & $ {\df{8 \sqrt2}{3}}(C_{23}-C_{12})$\\
%
 & $ \df{2}{3} (4C_{12}+14C_{23}+C_{13}+C_{24})$ \\
\end{tabular} \right]
\label{1++}
\end{equation}
which has been previously related  to X(3872). 
Its lowest state gave a mass of
3910 MeV to X(3872)\ \cite{Hogaasen:2005jv,Stancu:2006st}, quite close to the 
experimental value \ \cite{Choi:2003ue}.

For $J^P=1^{+-}$ there are four linearly independent basis 
vectors built as products of color states $b)$ 
and the first and second spin states of Eq. (\ref{directcolor1}) 
\begin{eqnarray}\label{eq:alphaprimi}
\hskip -10pt& \psi^1_{1^{+-}} =
| 1_{13} 1_{24} (P_{13} V_{24})_1 \rangle,\ 
&  \psi^2_{1^{+-}}  =
%
| 1_{13} 1_{24} (V_{13} P_{24})_1 \rangle,\nonumber\\
&\psi^3_{1^{+-}} =
%
| 8_{13} 8_{24} (P_{13} V_{24})_1 \rangle, 
& \psi^4_{1^{+-}} = 
| 8_{13} 8_{24} (V_{13} P_{24})_1 \rangle. 
\end{eqnarray}
The matrix associated to the chromomagnetic interaction -$H_{\mathrm{CM}}$ is
\\  

\vskip -20pt
\begin{equation}\left[
\renewcommand{\arraystretch}{2.0}
\begin {tabular}{cccccc}
%
$16(C_{13}-\df{1}{3}C_{24})$&0&0&$ 8 \sqrt{\df{2}{3}}(C_{12}+C_{23})$\\
%
0 &$-\df{16}{3}(C_{13}-C_{24})$&$ ~~~8 \sqrt{\df{2}{3}}(C_{12}+C_{23})$ &0\\
%
%
%
 & &$-2(C_{13}-\df{1}{3}C_{24})$ & $-\df{4}{3}(2C_{12}-7C_{23})$\\
%
%
 & & & $\df{2}{3}(C_{13}-3C_{24})$
\\
\end{tabular}\right]
\label{1+-}
\end{equation}

\vspace{3mm}
For $J^{PC} = 2^{++}$ the basis vectors are
\begin{eqnarray}\label{eq:deltai}
\hskip -10pt& \psi^1_{2^{++}} =
| 1_{13} 1_{24} \chi^S \rangle, \ 
&  \psi^2_{2^{++}} =
| 8_{13} 8_{24} \chi^S \rangle, 
\end{eqnarray}
where $\chi^S$ is the $S = 2$ spin state (\ref{SPIN2}).
The corresponding -$H_{\mathrm{CM}}$ 2 $\times$ 2 matrix is
\\
\vskip -20pt
\begin{equation}\left[
\renewcommand{\arraystretch}{2.0}
\begin {tabular}{cc}
%
$ -\df{16}{3}(C_{13}+C_{24})$ & $ {- \df{8 \sqrt2}{3}}(C_{23}-C_{12})$\\
%
 & $ - \df{2}{3} (4C_{12}+14C_{23}-C_{13}-C_{24})$ \\
\end{tabular} \right]
\label{2++}
\end{equation}
\vspace{3mm}
In the calculation of the matrix elements we have used the equalities
\begin{equation} 
C_{14} = C_{23}, ~~~C_{12} = C_{34},
\end{equation}
due to charge conjugation. 

The above matrices have been first used to calculate the full
spectrum of $c \bar c q \bar q$ with  $q = u,d$\ \cite{Stancu:2006st}. 
They  can be used in any quark model containing a chromomagnetic interaction.
In that case the coefficients $C_{ij}$ should be replaced by integrals
containing the chosen form factor of the chromomagnetic interaction and the
orbital wave functions of the model.

Note that the matrices (\ref{0++}), (\ref{1++}) and (\ref{2++})
have in common the off-diagonal matrix element $C_{23} -C_{12}$.
With the identification at the end of Sec. \ref{BS} this
leads to $C_{23} -C_{12} \equiv C_{c \bar s} - C_{cs}$. As $C_{c \bar s}$
and $C_{cs}$ have comparable values (\ref{eq:par}) their difference
is small.
In the next section 
we shall see that this off-diagonal matrix element plays an important role 
in the structure of the eigenstates with $J^{PC} = 0^{++}, 1^{++}$ 
and $2^{++}$.

\vspace{1cm}
\begin{figure}[h!]
\label{fig2}
\includegraphics*[width=10.0cm,keepaspectratio]{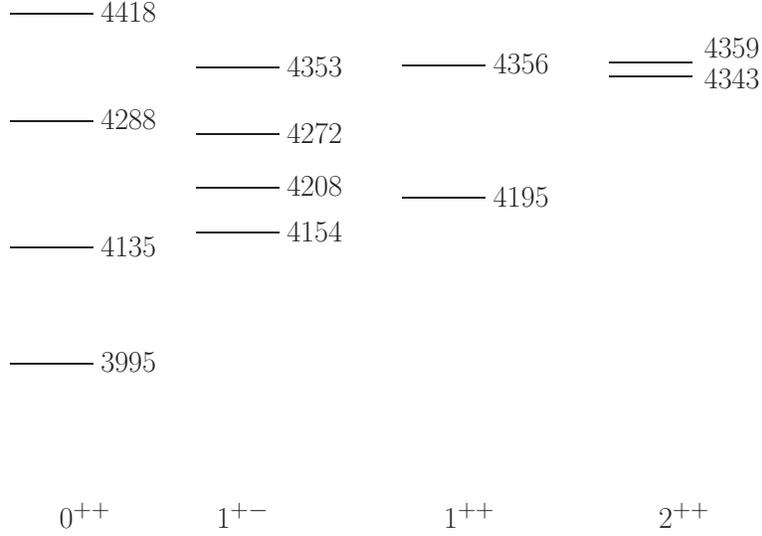}
\caption{The spectrum of the $c \bar c s \bar s$ tetraquark with  
the Hamiltonian introduced in Sec. \ref{HAM} and the color-spin bases of Sec. \ref{BS} }
\end{figure}

\section{The spectrum of $c \bar c s \bar s$}\label{SPECTRUM}

The calculated spectrum is exhibited in Fig. 2. 
There are several 
states in the range 4.1 - 4.2 MeV,
consistent with predictions of more realistic models \cite{Ebert:2005nc}.
This implies that the choice of the effective masses (\ref{eq:mass})
is quite adequate for $c \bar c s \bar s$ tetraquarks.
Here we are mostly interested in those states
with a small amplitude in the VV channel in the present parametrization.


\subsection{J$^{PC}$ = 0$^{++}$}

In the order indicated by the basis (\ref{eq:betai}) the lowest state, 
3995 MeV, has the amplitudes
\begin{equation}\label{lowest0++}
(-0.7737,~ 0.0594,~ 0.1789,~ 0.6049)
\end{equation}
The first number in the bracket implies that this state can decay substantially 
into a PP channel, \emph{i. e.} $\eta_c + \eta$ (threshold 3528 MeV) or 
$\eta_c + \eta'$ (threshold 3938 MeV). 
The second number indicates a very weak coupling to the  VV 
channel. The mass is too low for the decay into $J/\psi \phi$.

A better candidate would be the first excited state at 4135 MeV with
the amplitudes 
\begin{equation}\label{first0++}
(-0.6172,~ -0.1774,~ 0.4006,~ -0.6536)
\end{equation}
decaying substantially into PP channels and much less into the  VV 
channel  $J/\psi \phi$.
The last two amplitudes correspond  to hidden color channels
which do not decay strongly.

The tetraquark states mentioned above can also decay into the $D^+_s D^-_s$,
the threshold being 3936.68 MeV. 
The corresponding
amplitudes can be obtained from the orthogonal transformation going from
the direct meson-meson  channel, Fig. 1b, 
to the exchange meson-meson  channel, Fig. 1c. In Appendix \ref{appa}
we present the exchange basis vectors (\ref{eq:gammai}) in terms of
the direct basis vectors (\ref{eq:betai}) given by this transformation.
Using the expressions(\ref{ex1})-(\ref{ex4})
and the amplitudes 
(\ref{lowest0++}) obtained in the direct channel 
we can write the lowest state $0^{++}$ 
in the exchange channel basis as  
\begin{eqnarray}\label{EXCH}
\psi_{0^{++}} (3995) = -0.7244~ |1_{14}1_{23} \rangle |P_{14}P_{23} \rangle
+ 0.0743~|1_{14}1_{23} \rangle |(V_{14}V_{23})_0 \rangle \nonumber \\
- 0.2088~ |8_{14}8_{23} \rangle |P_{14}P_{23}\rangle 
+ 0.6529~ |8_{14}8_{23} \rangle |(V_{14}V_{23})_0 \rangle
\end{eqnarray}
Looking at Fig. 1c, again with  
1 = $c$, 2 = $s$, 3 = $\overline c$ and 4 = $\overline s$, 
we can identify  the color singlet-singlet channels in (\ref{eq:gammai})
with the
asymptotic meson-meson channels. Thus we have
\begin{equation}
\psi^{1ex}_{0^{++}} = |1_{14}1_{23} \rangle |P_{14}P_{23} \rangle = 
D_s \overline{D}_s
\end{equation}
\begin{equation}\label{D*D*}
\psi^{2ex}_{0^{++}} =  |1_{14}1_{23} \rangle |(V_{14}V_{23})_0 \rangle =
 D^*_s \overline{D}^*_s
\end{equation}
From the wave function (\ref{EXCH}) we can see that the open
$D_s \overline{D}_s$  channel acquires a large amplitude in the ground 
state (3995 MeV), corresponding to a probability of about 50\%,  which will
imply a large decay width in this channel and a negligible amplitude 0.5\%
to the closed channel $D^*_s \overline{D}^*_s$.


\subsection{J$^{PC}$ = 1$^{+-}$} 
In this sector the lowest state has an appropriate mass, but not the 
convenient charge conjugation. It would decay exclusively into a PV channel 
for $\ell$ = 0 tetraquarks. For general interest the exchange meson-meson
basis is also given in Appendix \ref{appb}.

\subsection{J$^{PC}$ = 1$^{++}$} 

For the lowest 1$^{++}$ state at 4195 MeV, which
is quite  close to the experimental range, the amplitudes of its  
components  in the
basis (\ref{eq:alphai}) are shown in Table \ref{one++} 
for two values 
of $C_{c \bar s}$.
\begin{table} 
\caption{The $D^*_s - D_s$ splitting (MeV) and the amplitudes of the
basis vectors (\ref{eq:alphai}) of the  
1$^{++}$  state at 4195 MeV  as a function of $C_{c \bar s}$ (MeV). }
\label{one++}
\renewcommand{\arraystretch}{1.8}
\begin{tabular}{ccccccc}
\hline 
 $C_{c \bar s}$   &\hspace{1cm}  & $D^*_s - D_s$ &\hspace{1cm} & 
 $1_{13}1_{24}(V_{13} V_{24})_1$ 
 & \hspace{1cm} &  $8_{13}8_{24}(V_{13} V_{24})_1$ \\  
\hline
6.0  & \hspace{1cm} & 128.0 & \hspace{1cm} & 0.0245 & \hspace{1cm} & 0.9997 \\
6.7  & \hspace{1cm} & 143.7 & \hspace{1cm} & 0.0399 & \hspace{1cm} & 0.9992 \\
\hline
\end{tabular}
\end{table} 
The singlet-singlet channnel $1_{13}1_{24}(V_{13} V_{24})_1$
has a very small amplitude for both values of $C_{c \bar s}$. 
The hidden color state  $8_{13}8_{24}(V_{13} V_{24})_1$ is  
by far the dominant component.  The situation 
is entirely analogous to that of the resonance
X(3872) 
in the same model \cite{Hogaasen:2005jv,Stancu:2006st}.
The clue was to have a nonvanishing, but small,  value for $C_{23}-C_{12}
\equiv C_{c \bar q}-C_{cq} $ in Eq.
(\ref{1++}). For X(3872) one had 1.5 MeV, here we have 
$ C_{c \bar s}-C_{cs}$ = 1.7 MeV
imposed by the parametrization (\ref{eq:par}). As
seen from Table \ref{one++} a decrease of $ C_{c \bar s}$
will make the hidden color state even more dominant 
but it will somewhat deteriorate the value of the $M_{D^*_s}- M_{D_s}$ 
splitting,
the experimental value of which is 143.8 MeV. 
Combined with the phase space of the decay Y(4140) $\rightarrow J/\psi \phi$
obtained from the experimental threshold, 
the lowest state 1$^{++}$ of the tetraquark $c \bar c s \bar s$ would acquire a rather 
small width, as required by experiment, and could be the best candidate
for Y(4140) in a tetraquark interpretation.

It is useful to write the wave function of the lowest state also in 
the exchange channel basis, as for the scalar tetraquaks above.
For this purpose we use the transformation between the direct and exchange
channel basis vectors derived in Appendix \ref{appb}, namely the
Eqs. (\ref{psi1ex}) and (\ref{psi2ex}).
Together with the amplitudes from  Table  \ref{one++}
associated to $C_{c \bar s}$ = 6.7 Mev we obtain for the lowest
${1^{++}}$ state
\begin{equation}\label{psi1++1ex}
\psi_{1^{++}}(4195) = - 0.9554~ \psi^{1ex}_{1^{++}} 
+ 0.2954  \psi^{2ex}_{1^{++}}.
\end{equation}
From Fig 1c and Eq. (\ref{eq:exch1+}) one has
\begin{equation}
\psi^{1ex}_{1^{+}} = D_s \overline{D}^{*}_s,
~~~ \psi^{2ex}_{1^{+}} = D^{*}_s \overline{D}_s 
\end{equation}
According to (\ref{psi_1ex_1++})  
a molecular-type component with $C = +$ is obtained in the exchange channel as
\begin{equation}
\psi^{1ex}_{1^{++}}  = \frac{1}{\sqrt{2}} (D_s \overline{D}^{*}_s - 
D^{*}_s \overline{D}_s )
\end{equation}
having a very large probability of 91.3 \% in the ${1^{++}}$ ground state. 
The phase space is larger than for the $J/\Psi \phi$ channel, so that a
large width is expected in the $D_s \overline{D}^{*}_s$ channel. 
The second term in (\ref{psi1++1ex}) is a hidden color component,
which does not decay, but vanishes asymptotically.


\subsection{J$^{PC}$ = 2$^{++}$} 
The spectrum is formed of two, nearly degenerate states,
both too high for Y(4140), by about 200 MeV. 
In the parametrization (\ref{eq:par})
the wave function
of the lowest state has the amplitudes 
\begin{equation}\label{lowest2++}
( - 0.4675,~0.8840)
\end{equation}
in the order of the basis  (\ref{eq:deltai}). 
One can see that the color singlet-singlet state 
$\psi^1_{2^{++}} =
| 1_{13} 1_{24} \chi^S \rangle$ 
has a small amplitude and
the hidden color state 
$\psi^2_{2^{++}} =
| 8_{13} 8_{24} \chi^S \rangle$ is dominant, again due to the smallness of
the off-diagonal matrix element $C_{c \bar s}-C_{cs} $. 
With $C_{c \bar s}$ = 6.0 MeV the amplitudes become
( - 0.2274,~0.9738). This would give rise to a even smaller decay width
into $J/\psi \phi$. 
The calculated mass fits better into the newly found narrow structure 
X(4350) reported by the Belle Collaboration \ \cite{Yuan:2009iu}. 
According to Appendix \ref{appc} the wave function
of the lowest state obtained from the latter amplitudes  
becomes 
\begin{equation}
\psi_{2^{++}}(4343)  = 0.8442~ {D}^*_s \overline{D}^{*}_s  
- 0.5390~ \psi^{2ex}_{2^{++}} 
\end{equation}
where we have replaced $\psi^{1ex}_{2^{++}}$ by its physical content. 
This state has a dominant molecular-type structure plus a hidden color 
component (\ref{psi2++2ex})
which would vanish asymptotically, but is important at short range.
In a standard hadronic molecule interpretation
\cite{Liu:2009ei,Mahajan:2009pj,Ding:2009vd,Branz:2009yt} the second
component is absent because the emitted mesons do not
have a structure.

\section{Conclusions}

Prior to the CDF experiment \cite{Aaltonen:2009tz}, among other
multiquark systems, the tetraquark
$c \bar c s \bar s$ has been studied with a different parametrization
from the one considered here and with a different basis using an SU(6) 
classification \cite{Cui:2006mp}.
In that basis it is difficult to identify the VV component.
Moreover a distinction between charge conjugation C = 1 and C = -1
has not been made. 

Our study favours mostly the 1$^{++}$ sector for the $c \bar c s \bar s$ 
tetraquark interpretation of the recently observed narrow structure 
Y(4140) \cite{Aaltonen:2009tz}. If correct,
Y(4140) would be the strange analogue of X(3872), when interpreted
as a $c \bar c q \bar q$ tetraquark. 
This observation follows from the fact that in the schematic model
of Ref. \cite{Hogaasen:2005jv} the chromomagnetic interaction leads
to a similar composition of the wave function in the basis 
(\ref{eq:alphai})
for tetraquarks containing either $u$ and/or $d$, like X(3872), or $s$ quarks,
like Y(4140). 

Note however that one should consider the effective masses (\ref{eq:mass})
with caution.
They have been obtained from fitting baryon and meson spectra. 
A natural question raises whether or not these masses are adequate for 
tetraquarks. This study questions their use in tetraquarks,  
inasmuch as they contain 
the effect of the kinetic energy and of the confinement.
The confinement has been thoroughly studied in lattice calculations.
A Y-shape confinement potential is
almost confirmed by  lattice results (see \emph{i. e.} 
 \cite{Takahashi:2000te}).
Information from lattice calculations on tetraquark 
(see \emph{i. e.} \cite{Okiharu:2004ve})
may lead to a better understanding of the effective masses
to be used in simple models. Thus, with the present parametrization
it is meaningless to look at the tetraquark spectrum relative
to the theoretical threshold. A detailed discussion based on a
simple example is presented in Appendix D.

Finally, we should mention that the present study does not exclude 
the 0$^{++}$ sector.
In fact, in the molecular $D^*_s {\overline D}^*_s$, 
Y(4140) can have the
quantum numbers 0$^{++}$ or 2$^{++}$.
As mentioned in the introduction, here we  stress again that  
the Belle Collaboration measurement of a two-photon partial
width  disfavors the scenario of Y(4140) to be a
$D^*_s \overline{D}^*_s$ molecule  with $J^{PC} = 0^{++}$
or $2^{++}$ \ \cite{Yuan:2009iu}.

A correct interpretation of the narrow structure Y(4140) observed by 
CDF  \cite{Aaltonen:2009tz}  would be
possible if its existence was confirmed
and its quantum numbers J$^{PC}$
were found experimentally, in order to remove the doubt cast 
by some theoretical interpretations \ \cite{vanBeveren:2009dc}. 
Also, the measurement of the decay widths
of other open channels such as 
$\eta_c + \eta$ or  $\eta_c + \eta'$ is important. If such decays are observed, 
the 0$^{++}$ sector 
is favored, if not, the sector 1$^{++}$ is favoured 
in a tetraquark interpretation. Complementary information can also be 
obtained from the decays to 
$D_s {\overline D}_s$, $D_s {\overline D}^{*}_s$, 
$D^*_s {\overline D}^{*}_s$  etc.

In conclusion, as a start, we have presented results in a tetraquark schematic model 
to get a hint on the interpretation of Y(4140), which remains an open problem.
Perhaps a more realistic view, if Y(4140) was confirmed,  
would be to have a compact 
tetraquark structure at short range and a molecular structure at medium 
or large range. Anyhow, a more elaborate study of the $c \bar c s \bar s$ tetraquark 
system is worth by itself. 
\section{Acknowledgments} 
I am most grateful to S. L. Olsen and S. Eidelman for valuable   
correspondence on the Belle Collaboration preliminary measurements.

\appendix

\section{Direct to exchange channel basis for $J^{PC} = 0^{++}$}
\label{appa}

In the following we need to express the color exchange basis in terms of 
the color direct basis vectors. The well known relations are 
\begin{equation}\label{colorex} 
|1_{14} 1_{23} \rangle = \frac{1}{3}|1_{13} 1_{24} \rangle 
+  \frac{2 \sqrt{2}}{3}|8_{13} 1_{24} \rangle,  \nonumber \\
~~~|8_{14} 8_{23} \rangle = \frac{2 \sqrt{2}}{3}|1_{13} 1_{24} \rangle 
-  \frac{1}{3} |8_{13} 1_{24} \rangle,
\end{equation}
Next, using Appendix C of Ref. \cite{Brink:1998as} for the spin states
we  obtain the spin-color exchange channel basis  in terms of the 
spin-color direct channel  basis (\ref{eq:betai}). 

For $J^{PC} = 0^{++}$ the exchange channel basis vectors are defined by
\begin{eqnarray}\label{eq:gammai}
\hskip -10pt& \psi^{1ex}_{0^{++}} = 
| 1_{14} 1_{23} P_{14} P_{23} \rangle,\
&  \psi^{2ex}_{0^{++}} =
| 1_{14} 1_{23} (V_{14} V_{23})_0 \rangle, \nonumber\\
\hskip -10pt& \psi^{3ex}_{0^{++}} =
| 8_{14} 8_{23} P_{14} P_{23} \rangle,\ 
&\psi^{4ex}_{0^{++}} =
| 8_{14} 8_{23}  (V_{14} V_{23})_0 \rangle .
%
%
\end{eqnarray}
In terms of the direct channel basis vectors (\ref{eq:betai}) the 
orthogonal transformation is given by the following relations
\begin{equation}\label{ex1}
\psi^{1ex}_{0^{++}} 
= \frac{1}{6}~ \psi^{1}_{0^{++}} - \frac{1}{2 \sqrt{3}}~ \psi^{2}_{0^{++}}
+ \frac{\sqrt{2}}{3}~ \psi^{3}_{0^{++}} - \sqrt{\frac{2}{3}}~ \psi^{4}_{0^{++}}
\end{equation}
\begin{equation}\label{ex2}
\psi^{2ex}_{0^{++}} 
= - \frac{1}{2 \sqrt{3}}~ \psi^{1}_{0^{++}} - \frac{1}{6}~ \psi^{2}_{0^{++}}
- \sqrt{\frac{2}{3}}~ \psi^{3}_{0^{++}} - \frac{\sqrt{2}}{3}~ \psi^{4}_{0^{++}}
\end{equation}
\begin{equation}\label{ex3}
\psi^{3ex}_{0^{++}} 
= \frac{\sqrt{2}}{3}~ \psi^{1}_{0^{++}} - \sqrt{\frac{2}{3}}~ \psi^{2}_{0^{++}}
- \frac{1}{6}~ \psi^{3}_{0^{++}} + \frac{1}{2 \sqrt{3}}~ \psi^{4}_{0^{++}}
\end{equation}
\begin{equation}\label{ex4}
\psi^{4ex}_{0^{++}} 
= - \sqrt{\frac{2}{3}}~ \psi^{1}_{0^{++}} - \frac{\sqrt{2}}{3}~ \psi^{2}_{0^{++}}
+ \frac{1}{2 \sqrt{3}} ~ \psi^{3}_{0^{++}} + \frac{1}{6}~ \psi^{4}_{0^{++}}
\end{equation}
These relations are used to derive Eq. (\ref{EXCH}).

\section{Direct to exchange channel basis for $J^{PC} = 1^{++}$}
\label{appb}

In the exchange channel corresponding to Fig. 1c the basis states
can be defined as above. Note however that in this case they do not
all have  a definite charge conjugation.  
Let us first introduce the 
$J^{PC} = 1^{+}$ the exchange channel basis vectors as 
\begin{eqnarray}\label{eq:exch1+}
\hskip -10pt& \psi^{1ex}_{1^{+}} =
| 1_{14} 1_{23} (P_{14} V_{23})_1 \rangle,\ 
&  \psi^{2ex}_{1^{+}} =
| 1_{14} 1_{23} (V_{14} P_{23})_1 \rangle,\nonumber\\
\hskip -10pt& \psi^{3ex}_{1^{+}} =
| 1_{14} 1_{23}~ (V_{14} V_{23})_1 \rangle, \  
& \psi^{4ex}_{1^{+}} =
| 8_{14} 8_{23} (P_{14} V_{23})_1 \rangle, \nonumber\\
\hskip -10pt&  \psi^{5ex}_{1^{+}}   =
| 8_{14} 8_{23} (V_{14} P_{23})_1 \rangle, \ 
& \psi^{6ex}_{1^{+}}  =
| 8_{14} 8_{23}~ (V_{14} V_{23})_1 \rangle.
\end{eqnarray}
Using Appendix C of Ref. \cite{Brink:1998as} which gives the
transformations in the spin space,  the exchange channel basis vectors  
(\ref{eq:exch1+}) can be written as linear combinations  of the direct 
channel basis vectors (\ref{eq:alphai}) and (\ref{eq:alphaprimi}). The
orthogonal transformation gives the equations

\begin{equation}
\psi^{1ex}_{1^{+}}= \frac{1}{6} (\psi^1_{1^{+-}} + \psi^2_{1^{+-}})
 - \frac{1}{3\sqrt{2}}~ \psi^1_{1^{++}}
+ \frac{\sqrt{2}}{3}(\psi^3_{1^{+-}} + \psi^4_{1^{+-}})
 - \frac{2}{3}~ \psi^2_{1^{++}}~ ,
\end{equation} 
\begin{equation}
\psi^{2ex}_{1^{+}} = \frac{1}{6} (\psi^1_{1^{+-}} + \psi^2_{1^{+-}})
+ \frac{1}{3\sqrt{2}}~ \psi^1_{1^{++}} 
+ \frac{\sqrt{2}}{3}(\psi^3_{1^{+-}} + \psi^4_{1^{+-}}) 
+ \frac{2}{3}~ \psi^2_{1^{++}}~ ,
\end{equation}
\begin{equation}
\psi^{3ex}_{1^{+}} = - \frac{1}{3\sqrt{2}} (\psi^1_{1^{+-}} - \psi^2_{1^{+-}}) 
- \frac{2}{3} ( \psi^3_{1^{+-}} - \psi^4_{1^{+-}})~,
\end{equation}
\begin{equation}
\psi^{4ex}_{1^{+}} = \frac{\sqrt{2}}{3} (\psi^1_{1^{+-}} + \psi^2_{1^{+-}})
- \frac{2}{3}~ \psi^1_{1^{++}}
- \frac{1}{6} (\psi^3_{1^{+-}} + \psi^4_{1^{+-}})
+ \frac{1}{3\sqrt{2}}~ \psi^2_{1^{++}}~,
\end{equation}
\begin{equation}
\psi^{5ex}_{1^{+}} = \frac{\sqrt{2}}{3}(\psi^1_{1^{+-}} + \psi^2_{1^{+-}})
+ \frac{2}{3}~ \psi^1_{1^{++}} 
- \frac{1}{6}(\psi^3_{1^{+-}} + \psi^4_{1^{+-}})
- \frac{1}{3\sqrt{2}}~ \psi^2_{1^{++}}~ ,
\end{equation}
\begin{equation}
\psi^{6ex}_{1^{+}} = - \frac{2}{3}(\psi^1_{1^{+-}} - \psi^2_{1^{+-}}) 
+ \frac{1}{3\sqrt{2}} ( \psi^3_{1^{+-}} - \psi^4_{1^{+-}})~.
\end{equation}
From these relations one can see that only $\psi^{3ex}_{1^{+}}$ and
$\psi^{6ex}_{1^{+}}$ have a definite charge conjugation $C = -$.
But in the exchange channel one can further introduce definite charge 
conjugation from the basis vectors in the following way.  
For $C = +$ the normalized states are
\begin{equation}\label{psi_1ex_1++} 
\psi^{1ex}_{1^{++}} = 
\frac{1}{\sqrt{2}}~ (\psi^{1ex}_{1^{+}} - \psi^{2ex}_{1^{+}})~,
\end{equation}
\begin{equation}\label{psi_2ex_1++}
\psi^{2ex}_{1^{++}} = 
\frac{1}{\sqrt{2}}~ (\psi^{4ex}_{1^{+}} - \psi^{5ex}_{1^{+}})~.
\end{equation}
Then for $C = -$ the normalized states are
\begin{equation}\label{psi_1ex_1+-} 
\psi^{1ex}_{1^{+-}} = \frac{1}{\sqrt{2}}~ (\psi^{1ex}_{1^{+}} + \psi^{2ex}_{1^{+}})~,
\end{equation}
\begin{equation}\label{psi_2ex_1+-} 
\psi^{2ex}_{1^{+-}} = \psi^{3ex}_{1^{+}}
\end{equation}
\begin{equation}\label{psi_3ex_1+-} 
\psi^{3ex}_{1^{+-}} = 
\frac{1}{\sqrt{2}}~ (\psi^{4ex}_{1^{+}} + \psi^{5ex}_{1^{+}})~,
\end{equation}
\begin{equation}\label{psi_4ex_1+-} 
\psi^{4ex}_{1^{+-}} = \psi^{6ex}_{1^{+}}~.
\end{equation}
Lastly, replacing the expressions of $\psi^{1ex}_{1^{+}}$, $\psi^{2ex}_{1^{+}}$,
$\psi^{4ex}_{1^{+}}$ and $\psi^{5ex}_{1^{+}}$
in  Eqs. (\ref{psi_1ex_1++}) and (\ref{psi_2ex_1++}) we get 
the orthogonal transformation relating the exchange channels with
the direct channel wave functions (\ref{eq:alphai}) for $C = +$
\begin{equation}\label{psi1ex}
\psi^{1ex}_{1^{++}} = - \frac{1}{3} \psi^{1}_{1^{++}} 
- \frac{2 \sqrt{2}}{3} \psi^{2}_{1^{++}} 
\end{equation}
\begin{equation}\label{psi2ex}
\psi^{2ex}_{1^{++}} = - \frac{2 \sqrt{2}}{3} \psi^{1}_{1^{++}} 
+ \frac{1}{3} \psi^{2}_{1^{++}} 
\end{equation}
This transformation will be used in the subsection C of Sec. V.

\section{Direct to exchange channel basis for $J^{PC} = 2^{++}$}
\label{appc}

The relations between the exchange and direct basis are in this 
case a direct consequence of the definitions (\ref{colorex})
inasmuch as the spin state $\chi^S$ is symmetric under any permutation
of $S_4$. One obtains 

\begin{equation}\label{psi2++1ex} 
\psi ^{1ex}_{2^{++}} = \frac{1}{3} \psi^{1}_{2^{++}}
+  \frac{2 \sqrt{2}}{3}  \psi^{2}_{2^{++}},
\end{equation}  
\begin{equation}\label{psi2++2ex}
\psi ^{2ex}_{2^{++}}= \frac{2 \sqrt{2}}{3}\psi^{1}_{2^{++}}
-  \frac{1}{3} \psi^{2}_{2^{++}}.
\end{equation}

This transformation will be used in the subsection D of Sec. V.
\section{Effective masses}
\label{effective}
First we establish the relation between  effective quark masses
used in these calculations and masses $m^0_i$ of a constituent quark model.  
For this purpose we start from the spin-indepenent part of a simple model 
of the commonly used type \cite{Bhaduri:1981pn}
\begin{eqnarray}\label{ham}
&&H_0= \sum_i m^0_i
  + \sum_i \frac{\vec{p}_{i}^{~2}}{2m^0_i}
  - \frac {(\sum_i \vec{p}_{i})^2}{2\sum_i m^0_i}
  + \sum_{i<j} \left[V_{\ell}(r_{ij})+ V_{C}(r_{ij})
  \right]\, 
\end{eqnarray}  
with a kinetic part from which the center of mass energy has been removed and
a potential part containing a two-body linear confinement 
$V_{\ell}(r_{ij})$ and a Coulomb-like term $V_{C}(r_{ij})$ 
\begin{eqnarray}\label{pot}
&&V_{\ell}(r_{ij}) = -\frac{3}{16}~\lambda_{i}^{c}\cdot\lambda_{j}^{c} \, 
~ ( \frac{r_{ij}}{a^2_0}- d) \, ,\quad
V_{C}(r_{ij}) =-\frac{3}{16}~\lambda_{i}^{c}\cdot\lambda_{j}^{c} \,
\frac{\kappa}{r_{ij}}. 
%
\end{eqnarray}
Together with a spin-spin part identical to that of
Ref. \cite{Bhaduri:1981pn} (not necessary to be specified here,
also used in other studies as \emph{e. g.}
Ref. \cite{Brink:1998as}), 
%
we have fitted the parameters of (\ref{ham}) to reproduce resonably well
the mass of $J/\psi$ and  $\phi$ mesons by choosing a trial wave function
of the form $\phi_0  
\propto \exp(-a^2 r^2_{ij}/2)$.
These calculations are aimed at understanding the basic reason behind the 
difference 
between effective masses and bare masses $m^0_i$. The fitted parameters are
\begin{eqnarray}\label{par}
&& m^0_c = 1600 \, {\rm MeV},~ m^0_s = 398 \, {\rm MeV},~
a_0 = 0.0361 \, {\rm MeV}^{-1/2} {\rm fm}^{1/2},\nonumber \\
&&~~~~~~~~~~~~~~d = 552.4 \, {\rm MeV}, ~\kappa = 39.47 \, {\rm MeV~ fm}.
\end{eqnarray}
\begin{table} 
\caption{Expectation values of $ H_0 $,  Eq. (\ref{ham}) and of its
kinetic and potential parts  
obtained from a trial wave function with the parameter $a$ (see text).}
\label{toy}
\renewcommand{\arraystretch}{1.8}
 \begin{tabular}{ccccccccc}
\hline 
System    &\hspace{1cm}  & $a$   & \hspace{1cm} & Kinetic  & \hspace{1cm} &
Potential & \hspace{1cm} & $\langle H_0 \rangle$   \\
          &\hspace{1cm}  & (fm$^{-1})$    & \hspace{1cm} & (MeV)      
	  & \hspace{1cm} & (MeV)         & \hspace{1cm} & (MeV) \\
\hline
$c \bar c $& \hspace{1cm} & $2.5$ & \hspace{1cm} & 229.6    & \hspace{1cm} &
 -317.8  & \hspace{1cm} &  3092 \\
$s \bar s $& \hspace{1cm} & $1.4$ & \hspace{1cm} & 336.7    & \hspace{1cm} & 
   3.27  & \hspace{1cm} & 1020 \\
$c \bar c s \bar s $   & \hspace{1cm} & $2.1$ & \hspace{1cm} & 656.8  & 
\hspace{1cm} & -39.8 &  \hspace{1cm} & 4477 \\
\hline
\end{tabular}
\end{table}
Below they are used to estimate the expectation value of $ H_0 $
corresponding to a $c \bar c s \bar s $ system described by a trial wave 
function of the form  $R \propto \exp[-a^2(\sigma^2+\sigma'{~^2}+\lambda^2)]$,
$a$  being a variational parameter, as above. Here, for convenience, we use
the internal coordinates
\begin{equation}\label{radial}
\vec{\sigma}  = \frac{1}{\sqrt{2}} (\vec{r_1}-\vec{r_2})\, ,\quad
\vec{\sigma'} = \frac{1}{\sqrt{2}} (\vec{r_3}-\vec{r_4})\, ,\quad
\vec{\lambda} = \frac{1}{2}(\vec{r_1}+\vec{r_2}-\vec{r_3}-\vec{r_4}).
\end{equation}
corresponding to Fig. 1a.
Note that this function  can be defined in any of the coordinate systems 
of Fig. 1.
Using the coordinates 
(\ref{radial})  we can work out the matrix elements of the flavor
operators of (\ref{pot}) in the basis  
$ ~|\overline{3}_{12} 3_{34} \rangle, ~ |6_{12} \overline{6}_{34} \rangle $
of Sec. III. 
The desired expectation values obtained with the parametres (\ref{par})
are shown in Table \ref{toy} for all considered systems.

From these results we can define effective masses in a similar way as 
in Ref. \cite{Hogaasen:2005jv}. We have 
\begin{equation}
m^{eff}_{q} = \frac{1}{2}\langle H_0 \rangle_{q \bar q}
\end{equation}
which lead to
\begin{equation}\label{eff:mass}
\begin{array}{lll}
 m^{eff}_{c}=1546\,\mathrm{MeV},&
m^{eff}_{s}=510\,\mathrm{MeV}.
\end{array}\end{equation}
Although we rely on the same PDG data \cite{Amsler:2008zzb}  
these masses are different from those of Eq. (\ref{eq:mass})
proposed in  Ref. \cite{Hogaasen:2005jv}. The difference is however 
very small for the $c$ quark and this can be explain by the cancellation of 
the kinetic and potential energies, as one can see from Table \ref{toy}.
Such a cancellation does not take place for the quark $s$. 
Thus in a dynamical approach based on a Hamiltonian like (\ref{ham})
there is a cancellation of various parts of the Hamiltonian. The  
cancellation is more subtle  
in a tetraquark which has 6 distinct quark-quark or quark-antiquark pairs, 
while in a meson there is only one
pair. It follows then that the effective masses 
needed for a tetraquak can be different from those of Eq. (\ref{eff:mass}).
Indeed, using Table \ref{toy}, we obtain 
\begin{equation}\label{tet}
m^{eff}_{c}+ m^{eff}_{s} = \frac{1}{2}\langle H_0 \rangle_{c \bar c s \bar s}
= 2238.5\,\mathrm{MeV}
\end{equation}
which is different from the sum of masses in (\ref{eff:mass}).
This proves that one cannot use the same effective masses in mesons and 
tetraquarks. In this light we can consider the choice (\ref{eq:mass})
acceptable and understand why the agreement with the experiment in Table 
\ref{pdgmass} is unsatisfactory for mesons. A better knowledge of the confinement 
and more precise calculations are necessary to obtain the mass of the
$c \bar c s \bar s $ tetraquark relative to the $J/\psi \phi$ threshold.


\end{document}